\documentclass[english]{sobraep}

\usepackage{ragged2e}

\title{The SLED (Shelf Life Expiration Date) Tracking System: Using Machine Learning Algorithms to Combat Food Waste and Food Borne Illnesses}

\author{Srilekha Mamidala\\
	\normalsize Garnet Valley, Pennsylvania, United States of America\\
	\normalsize E-mail: srilekha@wharton.upenn.edu
}

\begin{document}

\maketitle

\begin{abstract}
	The issue of food waste is a major problem contributing to the emission of greenhouse gases into the environment in addition to causing illness in humans. This research aimed to develop a correlation between the amount of time until a food spoiled and dates on food labels in conjunction with sensory observations. Sensory observations are more accurate as they are immediate observations that are specific to the food. This experiment observed bananas, bread, milk, eggs, and leafy greens over a period of time using characteristics specific to the food to quantify food spoilage. It was shown that the actual time until spoilage for all foods was longer than that of the best by date and that sensory observations proved to be a more accurate factor in determining spoilage. From this data, a machine learning algorithm was trained to predict if food was spoiled or not, in addition to the number of days until spoilage. This was presented to the consumer as an app, where the user can track foods and are reminded to check on them to prevent wastage. In addition, the experimental procedures were incorporated into a test kit for the consumer to take instructed observations to assess the spoilage of their food, which are then entered into the app to improve the algorithm. This paper discusses the individual effects of sensorial observations on each food and examines the shifting of consumer habits through an app and test kit to combat environmental consequences of food waste.

\end{abstract}


%~~~~~~~~~~~~~~~~~~~~~~~~~~~~~~~~~~~~~~~~
%Sections
%~~~~~~~~~~~~~~~~~~~~~~~~~~~~~~~~~~~~~~~~

%Introduction

\section{INTRODUCTION}

Every year, ambiguity surrounding the expiration of foods causes shoppers to waste 1.6 billion tons of food, harming the environment by releasing methane, a potent greenhouse gas. Additionally, 1 in 6 people get sick each year from consuming expired foods that have been exposed to dangerous bacteria. This project explored when certain foods truly spoil based on experimental observations instead of the misleading labels on food packaging. These labels provide guidelines on freshness, but many consumers prematurely toss out perfectly safe food thinking that they imply expiration. Consequently, these foods end up in landfills, wasting resources and damaging the environment.

The issue of food waste affects consumers around the world. To avoid misleading consumers, it is important to prove that the best by dates do not indicate spoilage, and that foods often last significantly longer than the best by dates on printed food labels. In addition, it is essential to develop a system in which the consumer can test and track the spoilage of their food using sensorial observations to promote sustainable consumer habits that help raise awareness and reduce the amount of food waste that makes it to landfills. 

The environmental effects of food waste are astounding. With many calling it a crisis, food waste is a major contributor to global warming, along with many other serious effects on society. In fact, if wasted food were classified as its own country, it would only be third in the emissions of carbon dioxide in the world, only behind the US and China [1]. 

Current devices to test spoilage are not catered to consumers and require specialized, expensive equipment that are only available in a laboratory setting. Applications serving as reminder systems to prevent wasted food are often catered to grocery stores and food companies to track food on a large scale [5]. The few reminder applications marketed to the consumer rely on the “best-by” dates on food labels, which are inaccurate when determining spoilage and should only be used as a guideline for freshness. 

This paper aims to explain the correlation between the true time it takes a food to spoil and sensorial observations in conjunction with best by dates.  It was hypothesized that the experimental amount of time for the foods to spoil would be greater than the date on the food label and that sensory observations would provide the most guidance on when foods are spoiled because dates on food labels are meant to indicate freshness, not spoilage. Additionally, sensory observations are more accurate as they are immediate observations that are specific to the food. The results of this experiment can be used to protect the environment by reducing methane emissions, in addition to saving resources that could potentially be used to feed people in need.

%Methods

\section{EXPERIMENTAL}

\begin{center}
    \includegraphics[width=0.45 \textwidth]{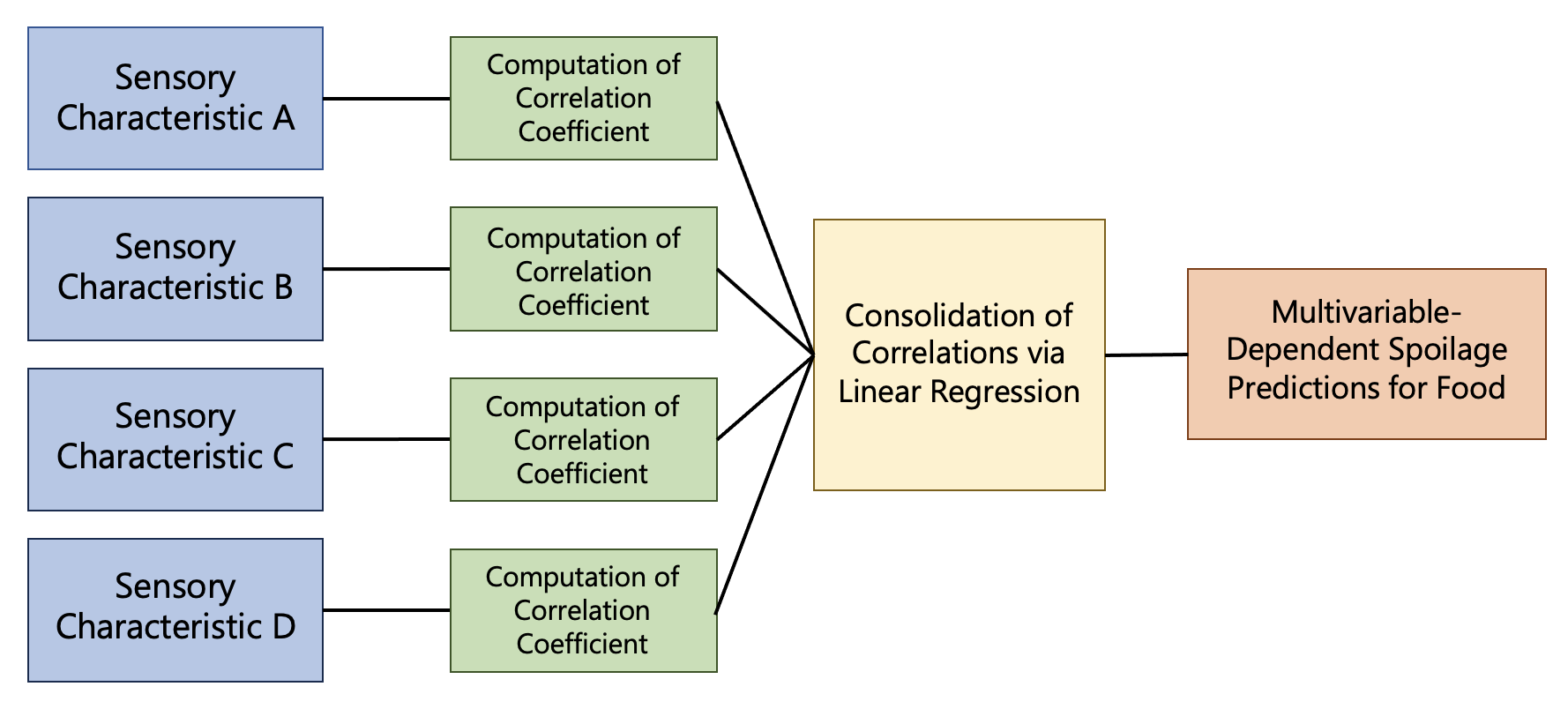}
\end{center}

\centering \emph{Figure 1: Proposed approach for quantification and modeling of food spoilage}

\justifying

\subsection{Foods Tested} 

This experiment had five trials and tested the spoilage of five foods that are known to be some of the most wasted foods: bread, milk, eggs, bananas, and leafy greens. Each food was assigned 3-7 characteristics that could be measured to determine if the food was spoiled or not. They were observed based on known estimates on how long these foods usually lasted, using characteristics specific to the food to determine when these foods spoiled. These characteristics are listed in the figure below.

\begin{center}
    \includegraphics[width=0.45 \textwidth]{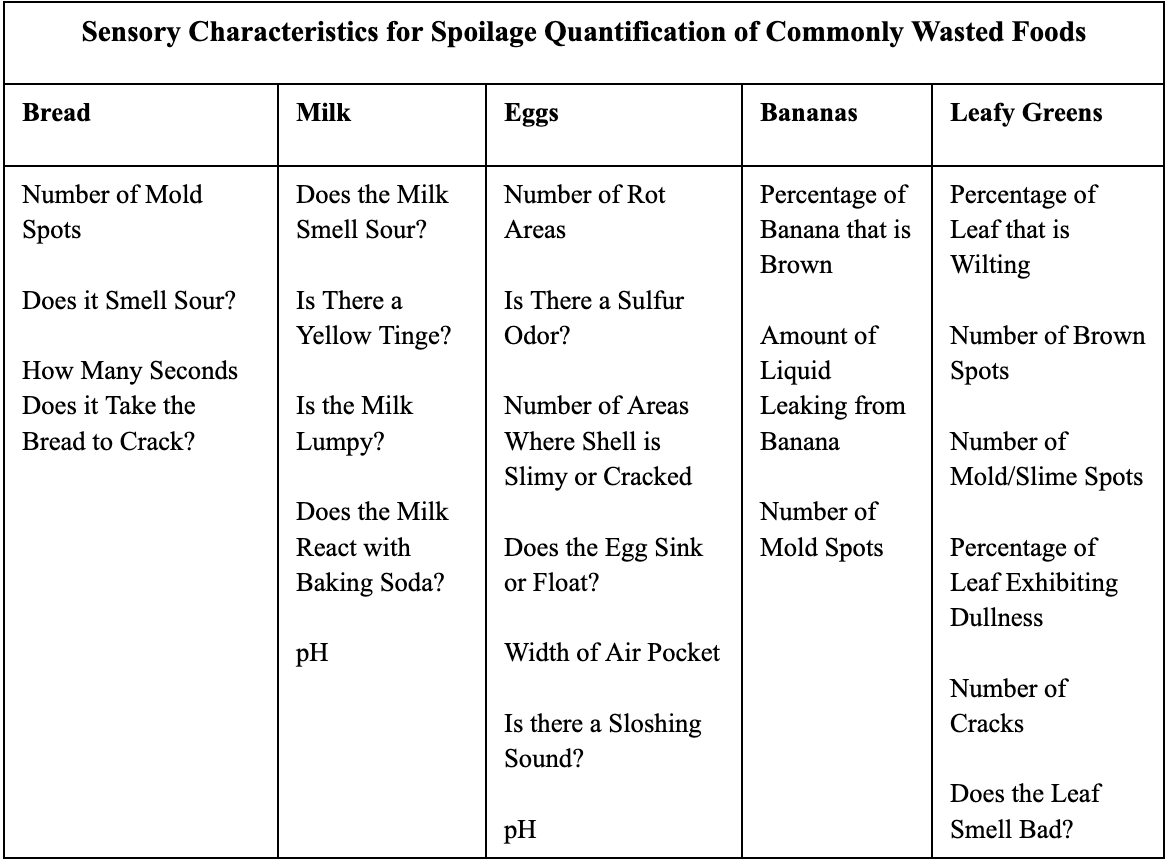}
\end{center}

\centering \emph{Figure 2: Table displaying characteristics observed during the experiment}

\justifying

\vspace{5mm} %5mm vertical space

The bread was observed for 18 days, the milk was observed for 50, the eggs for 60 days, the bananas for 12 days, and the leafy greens for 12 days as well. Each food was sourced from a well-known brand. Foods that were refrigerated, such as the milk, eggs, and leafy greens, were stored at 35ºF (1.7ºC). Foods tested in a room-temperature environment, such as the bread and bananas, had an ambient temperature of 73ºF (22.8ºC). 

\subsection{Procedures for Sensory Characteristics} 

Sensory details checking for the presence of a certain characteristic (ex. if the milk has a yellow tinge) or asking for a certain number (ex. number of mold spots) were verified by asking multiple people to observe the food. Additionally, photos were taken of every food at each time of observation and were reviewed after the experiment.
pH was measured immediately after it was dipped into the food.

To measure if the milk reacted with the baking soda, milk was extracted with a pipette and placed in baking soda [9]. Bubbling indicated spoilage, while no reaction did not.

Measuring the number of seconds a slice of bread took to crack measured staleness. This was done by applying constant pressure from a clothespin onto the bread and measuring the number of seconds until the bread cracked.

The width of the air sac of the egg was measured by placing the egg in a dark room and illuminating the bottom using a flashlight, revealing the air sac. The width was measured using a ruler, with a wider air sac indicating spoilage [2].

To test the amount of liquid leaking from the banana, a pipette was used to extract the liquid and measure it to check for the presence of liquid, as the presence itself indicates that the banana is unsafe to eat [3].

To check if the egg would sink or float, the egg was simply placed in a tub of water [6].

\begin{center}
    \includegraphics[width=0.45 \textwidth]{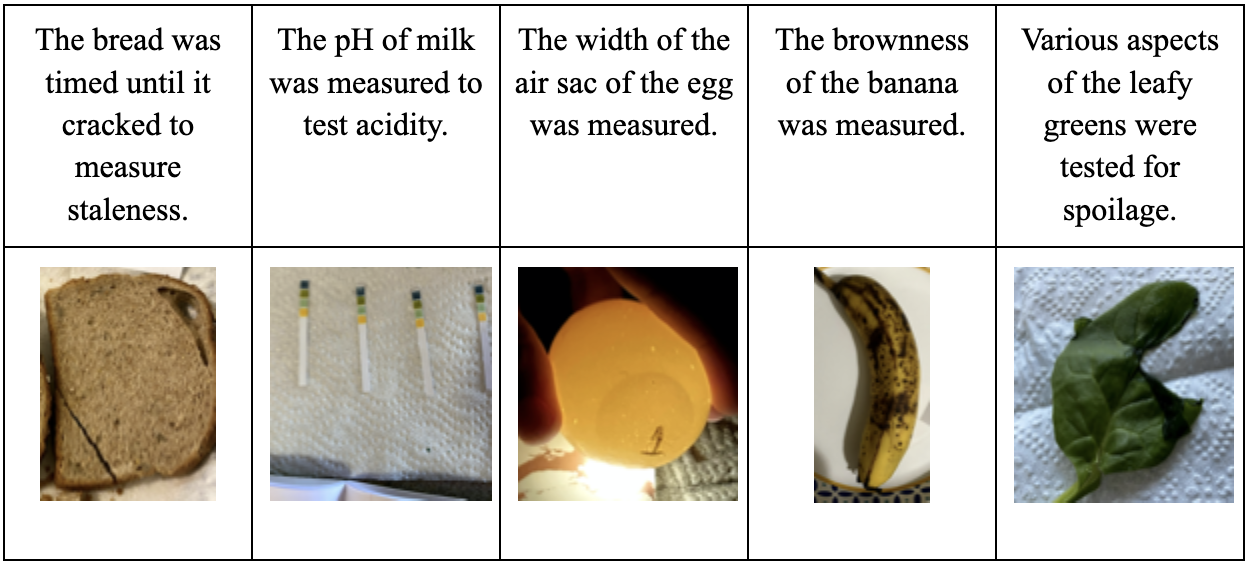}
\end{center}

\centering \emph{Figure 3: Shown above are some characteristics of the foods being tested.}

\justifying

\vspace{2mm} %5mm vertical space

\subsection{Linear Regression Analysis} 

After data was collected, a linear regression was calculated for each characteristic. Each of these equations were added for each food to create a “spoilage score” representing the total spoilage. To calculate a linear regression equation in the form of $y=a+bx$, Pearson's correlation coefficient was used to measure the correlation between the day they were observed and the value of the characteristic tested [7]. The procedure and the derived linear regression equations of each food is as follows. These were compiled into a data set that could be read into the machine learning algorithm to train it.

To calculate linear regression in the intended form, Pearson's correlation coefficient, $r$, was used to measure the correlation between the day and the value of the characteristic tested. 

Pearson's correlation coefficient, $r$, can be computed mathematically:

\begin{center}

 {\Large$r = \frac{\sum ((x-\overline{x})(y-\overline{y}))}{\sqrt{\sum (x-\overline{x})^2(y-\overline{y})^2}}$},
 
\end{center}

where:
\begin{itemize}
\item $x$ = value of characteristic measured
\item $y$ = day of characteristic measurement
\end{itemize}

The standard deviations can be computed as follows:

\begin{center}

 {\Large$r = \sqrt{\frac{\sum (x-\overline{x})^2} {n-1}}$}

 {\Large$r = \sqrt{\frac{\sum (y-\overline{y})^2} {n-1}}$}
\end{center}

where:
\begin{itemize}
\item $S_x$ = standard deviation of $x$-values
\item $S_x$ = standard deviation of $x$-values
\item $n$ = number of data points
\end{itemize}

Through these calculations, the slope of a regression line can be calculated:

\begin{center}

 {\Large$b = r*\frac{S_y}{S_x}$}.

\end{center}

The derived linear regression equations for the five foods observed were subsequently calculated:

\begin{center}

Banana: \\

{$8.142+1.333*$(\emph{Estimated Percentage of Browning in Banana}) $+ 3.345*$(\emph{mL of Liquid Leaking from Banana}) $+2.012*$(\emph{Number of Mold Spots on Banana})}.
\vspace{3mm} %5mm vertical space
Bread: \\

{$39.206 + 3.135*$(\emph{Number of Mold Spots}) $+ 19.070*$(\emph{Binary Classification of Sourness}) $+ 0.2977*$(\emph{Number of Seconds for Bread to Crack})}
\vspace{3mm} %5mm vertical space

Egg: \\

{$-69.663 + 11.287*$(\emph{Number of Rot Areas}) $+ 40.024*$(\emph{Binary Classification of Sulfur Odor}) $+ 0.086*$(\emph{Number of Slimy Areas })$+ 75.000*$(\emph{Binary Classification of Buoyancy (Sink/Float)})$+ 67.015*$(\emph{Width of Air Pocket})$+ 40.024*$(\emph{Binary Classification of Sloshing})$+ 20.757*$(\emph{pH of Egg White})}
\vspace{3mm} %5mm vertical space

Leafy Greens: \\

{$24.450 + 0.1554*$(\emph{Percentage of Greens Wilting}) $+ 0.5045*$(\emph{Number of Brown Spots on Leaves}) $+ 1.333*$(\emph{Number of Mold or Slime Spots}) $+ 0.1275*$(\emph{Percentage of Greens Exhibiting Dulled Coloring}) $+ 1.552*$(\emph{Number of Cracks in Leafy Greens}) $+ 8.000*$(\emph{Binary Classification of Smell})}
\vspace{3mm} %5mm vertical space

Milk: \\

{$323.404 + 30.000*$(\emph{Binary Classification of Sourness}) $+ 39.647*$(\emph{Binary Classification of Yellow Tinge}) $+ 30.743*$(\emph{Binary Classification of Presence of Lumps}) $+ 28.841*$(\emph{Result of Baking Soda Test}) $+ 37.712*$(\emph{pH of Milk})}

\end{center}

To present this algorithm in a user-friendly manner, an app was created that would allow the user to submit their observations. To determine if the food was spoiled or not,  a boundary value was set for the spoilage score. If the calculated spoilage score was higher than the boundary, the food was deemed spoiled, and if it was lower, the food was considered to be safe to eat. The boundary value was calculated from the regression graphs. When a food is spoiled, the majority of the sensory characteristics reflect this, shown by a considerable change in the spoilage scores of the regression graphs [5]. Each time data is inputted into the app, the regression is recomputed to account for the new data, making predictions more accurate over time. 

\section{RESULTS AND DISCUSSION}

\subsection{Modeling of Spoilage}

Following the consolidation of correlation coefficients to produce spoilage models for each food, the individual observations were computed via the model and were plotted via a linear regression. 

\subsubsection{Banana Spoilage}

\begin{center}
    \includegraphics[width=0.45 \textwidth]{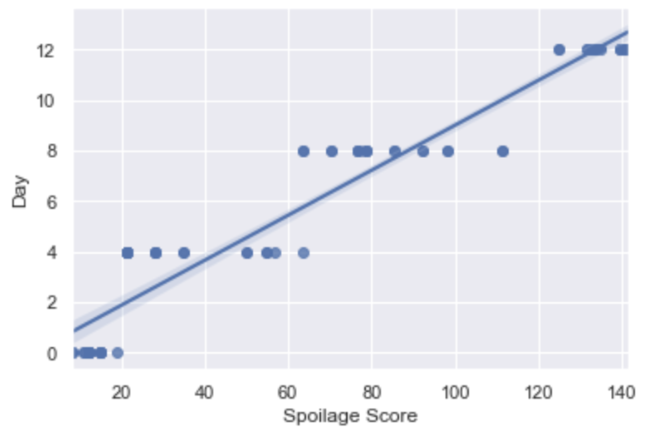}
\end{center}

\centering \emph{Figure 5: Spoilage of the banana is plotted to reveal a regression graph. Each individual point corresponds to the spoilage of the banana after a specific number of days.}

\justifying

\vspace{3mm} %5mm vertical space

Pearson’s correlation coefficient for the bananas was very strong for all characteristics. The graph showed a constant rate of spoilage of the bananas. The boundary spoilage score was placed at 50 because this score indicated the first presence of mold, indicating spoilage. This placed the true number of days a banana takes to spoil under the experimental conditions at 4-5 days. 

\subsubsection{Bread Spoilage}

\begin{center}
    \includegraphics[width=0.45 \textwidth]{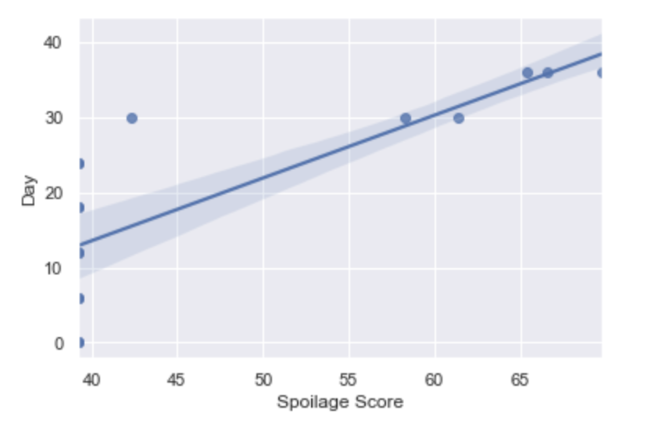}
\end{center}

\centering \emph{Figure 6: Spoilage of the bread revealed a regression graph. Each point corresponds to the spoilage of the bread after a specific number of days, computed by the spoilage score. The regression line was computed by the machine learning algorithm. }

\justifying

\vspace{3mm} %5mm vertical space

Correlation between the characteristics and the number of days was not as strong. However, the spoilage scores were considerably higher, indicating that the bread did not spoil at a constant rate. The boundary spoilage score was placed at 45 because the spoilage scores of bread approached but did not pass this value, placing the number of days bread takes to spoil at approximately 20-25 days.

\subsubsection{Egg Spoilage}

\begin{center}
    \includegraphics[width=0.45 \textwidth]{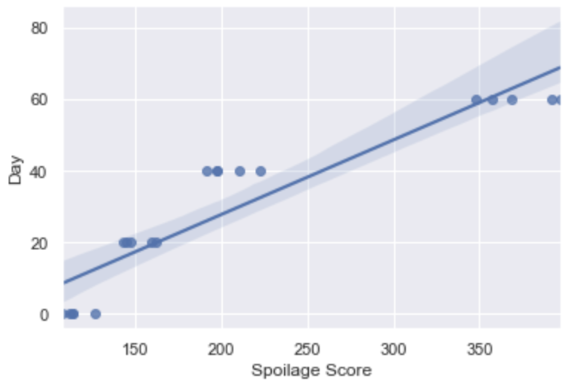}
\end{center}

\centering \emph{Figure 7: Spoilage of the egg was plotted, revealing a regression graph. Each individual point corresponds to the spoiling egg after a specific number of days, computed by the spoilage score.}

\justifying

\vspace{3mm} %5mm vertical space

Many of the characteristics tested supported a strong correlation with the day and supported the conclusion that eggs spoiled at a constant rate until a certain number of days, shown by the dramatic increase in spoilage scores after the 40 day mark. For this reason, the boundary spoilage score was calculated as 250 and the predicted number of days eggs take to spoil is 40.

\subsubsection{Milk Spoilage}

\begin{center}
    \includegraphics[width=0.45 \textwidth]{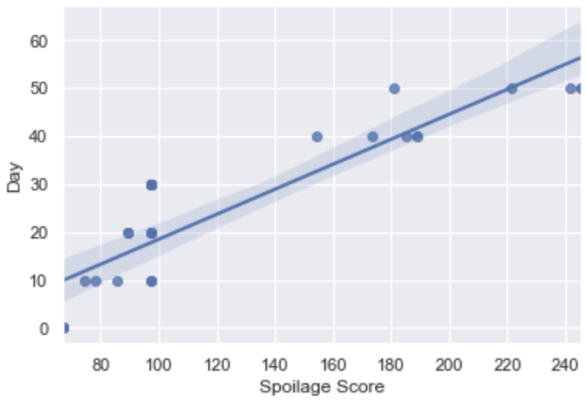}
\end{center}

\centering \emph{Figure 8: Spoilage of the milk revealed a regression graph. Each point corresponds to the spoilage of the milk after a specific number of days, computed by the spoilage score. The regression line was computed by the algorithm.}

\justifying

\vspace{3mm} %5mm vertical space

The graph shows that a significant increase in spoilage scores was detected after the 30 day mark, demonstrating that spoilage does not occur at a constant rate and only occurs after a certain amount of time. The boundary spoilage score was placed at the 140 mark, appointing the number of days milk takes to spoil at 30 days. 

\subsubsection{Leafy Greens Spoilage}

\begin{center}
    \includegraphics[width=0.45 \textwidth]{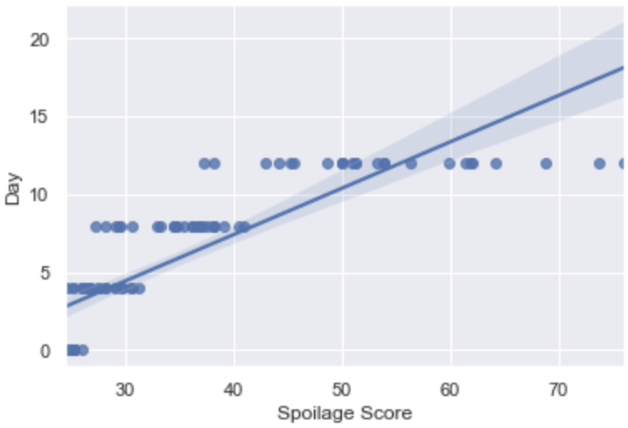}
\end{center}

\centering \emph{Figure 9: Spoilage of the leafy greens were plotted, revealing a regression graph. Each individual point corresponds to the spoilage of the leafy green after a specific number of days, computed by the spoilage score.}

\justifying

\vspace{3mm} %5mm vertical space

In the case of the leafy greens, Pearson’s correlation coefficient indicated a strong correlation between all characteristics and the day the characteristic was measured. The graph proves a constant rate of spoilage. However, past the 8 day mark, spoilage scores became more varied as leaves exhibited different degrees of spoilage, causing the boundary spoilage score to be placed at 40 to account for this variance.

\vspace{5mm} %5mm vertical space

Overall, this study found that the staple foods tested last longer than the respective dates printed on food labels. Compared to published research, the bananas took the same amount of time to spoil. The bread took 3-4 times longer to spoil than the length of time printed on the food label, aligning with previous research studies finding that bread was often wasted before spoiling. According to previous studies, eggs are expected to take 3-5 weeks to spoil. This experiment verified this and proved that eggs last at least two weeks after the sell by date. The milk lasted twice as long past the date printed on the sell by label. This was a longer amount of time than predicted by other sources which may be due to the storage environment or type of milk [4]. The leafy greens lasted approximately the same amount of time as published studies suggested, and several days past the best by date on the package. 

During this experiment, several questions that arose included which sensory observations to perform, as some were more influential than others. To determine which were the most important to include in the text kit, Pearson’s correlation coefficients were compared and the characteristics with the strongest correlation were selected. Additionally, several procedures had to be redefined to accurately reflect the conditions that food would be placed in. In pre trials of the bread, the bread was placed by itself, which caused staling to occur far faster than it would if the bread was placed in its customary plastic packaging. To mimic this, the bread was placed in plastic bags such that there was still some form of air flow.

Some errors that were present during the experiment included the collection of data. Although this error was minimized by having the observations be verified by multiple people, purely observational data is still subject to error and may have affected the machine learning algorithm. Additionally, in order for this experiment to deliver accurate results, the foods must be stored in similar conditions as testing. 

\subsection{SLED Application}

As aforementioned, to present this algorithm and the results of the experiment in a more accessible way, an app was created in which the user could input their observations and receive immediate feedback on whether their food was spoiled or not, in addition to how many more days their food could be expected to last if they were still safe to eat under the same conditions. In addition, the app also serves as a reminder system, alerting the user when their food is close to expiring so it can be consumed before it is unsafe and needs to be thrown away. The app was designed using Swift, the programming language for iOS apps. It also includes features for the user to control reminders, add foods, and set initial conditions for the foods to adjust the output from the algorithm, as shown in Figure 9. 

\begin{center}
    \includegraphics[width=0.48 \textwidth]{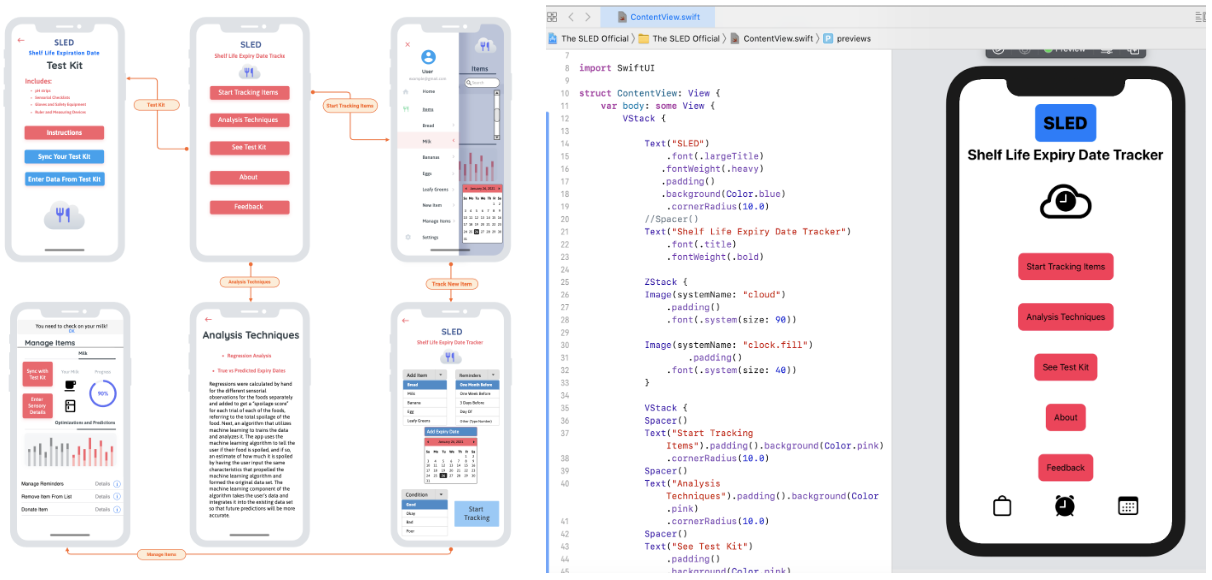}
\end{center}

\centering \emph{Figure 10: On the left, features of the app such as the tracking of foods and the presetting of initial conditions are shown. On the right, the GUI of the app was programmed using Swift.}

\justifying

\vspace{5mm} %5mm vertical space

\subsubsection{Test Kit Implementation}

Furthermore, the experimental procedures were incorporated into a test kit for the consumer to take instructed observations to assess the spoilage of their food, which are then entered into the app to improve the algorithm. A test kit for the user to test all of the characteristics that were used in the experiment was necessary because some required additional materials The test kit make the observation process more efficient by not requiring the user to find their own materials, it also serves as an alternative to people who may not have access to the technology required for the app. Food waste is a problem that affects consumers from all backgrounds. The addition of a test kit ensured that everyone could measure and check if their food is spoiled, with or without technology. Linking an app with the test kit promoted a system in which the consumer controls their own food, promoting consumer responsibility and helping lead to a change in consumer habits that is essential to solve the problem of food waste. 

Safety glasses were included to prevent contamination or illness from potentially spoiled foods. A thermometer was included to test the temperature of foods to test for spoilage at a certain temperature. pH strips were used to test the pH of the milk and eggs. The flashlight and ruler are included for the user to measure the width of the air sac of the egg in a dark room [2]. A pipette is needed to extract samples of the food to test certain characteristics. The test tube provides a separate vessel to hold the milk to make observations. A checklist is included as a guideline for the user to measure characteristics and to measure spoilage in the absence of the app. The test kit with these necessary materials is shown in Figure 10. 

\begin{center}
   \includegraphics[width=0.48 \textwidth]{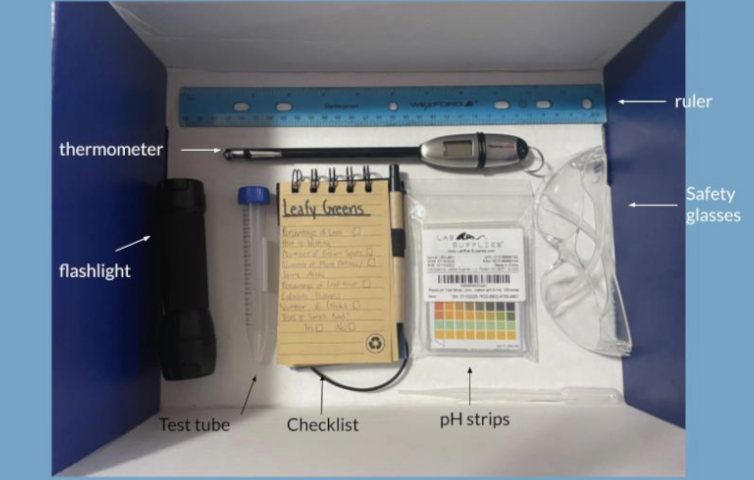}
\end{center}

\centering \emph{Figure 11: A test kit was created with necessary materials for the user to test the spoilage of their own foods, which can be inputted into the app.}

\justifying

\vspace{5mm} %5mm vertical space

\subsection{Impact and Applications}

\subsubsection{Environmental Sustainability of Resources}

Consuming food that would have otherwise been wasted saves millions of consumer dollars each year. Food waste also hurts the farmers that make the ingredients for these foods. Throwing out a single glass of milk may not seem to pose a significant problem. However, for every glass of milk thrown away, 1000 liters of water are wasted because this is the amount of water used in the process to pasteurize, feed the cows that supply the milk, and more. In addition, this glass of milk required large areas of land for these cows. As 1/3 of the world’s land dedicated to agriculture ends up growing wasted food, this land often replaces the habitats of wild animals, posing another environmental concern: habitat loss from human activity, especially deforestation [1]. This puts ecosystems in jeopardy and is the greatest threat to animal species. To get the milk to a grocery store, oil is needed to transport the milk and to drive production processes. Every year, millions of gallons of oil used in these processes are being wasted, contributing to the environmental issue of non-renewable resources, which emit a considerable amount of carbon dioxide into the atmosphere, further worsening the problem of global warming. Saving a single glass of milk, or any other food that could have been potentially wasted, can play a part, however small, in saving our planet.

Using the app would allow consumers to take control of their own food by tracking it and having established guidelines to know if their food is spoiled and, if applicable, when their food will spoil. Not only does this help save resources, it foundationally shifts consumer habits that are sustainable and necessary to solve this problem.

\subsubsection{Preventing Illness from Expired Foods}   

The app has a reminder feature that alerts the user to check on their food before it spoils. The purpose of this feature was to prevent food from being consumed after it was spoiled, which can lead to food poisoning that can potentially be fatal.

Food poisoning affects millions of consumers every year, often in their very own homes. Current food surveillance software is catered towards industrial suppliers or restaurants to monitor food spoilage to avoid serving spoiled food to consumers. However, the SLED addresses the problem of food induced illness on a exclusive consumer level and is the first of its kind to do so.

\justifying

\subsubsection{Reducing Greenhouse Gas Emissions}

Food waste in landfills produces methane, a greenhouse gas that has a heat trapping ability 21 times that of carbon dioxide. In addition, food waste contributes 135 million tons of greenhouse gas emissions each year, a major contributor to global warming and worsening the climate crisis the world is already facing. Using the app to check if food is spoiled will help consumers cut down on the amount of food that makes it to landfills every year, reducing the amount of methane that is produced.

\subsubsection{Establishment of Effective Legislation}

This project may help establish legislation concerning the use of best by labels to reduce confusion to the consumer. In addition, it may enact programs to provide food at risk of being wasted to vulnerable people. All of the food wasted every year is enough to feed 2 billion people, twice the number of malnourished people in the world today [8]. Cutting down on the amount of wasted food can lead to more food being donated to people in need that would have been otherwise thrown away. Additionally, to help raise awareness about the food waste crisis, a website was created, shown in Figure 11. 

\begin{center}
   \includegraphics[width=0.5 \textwidth]{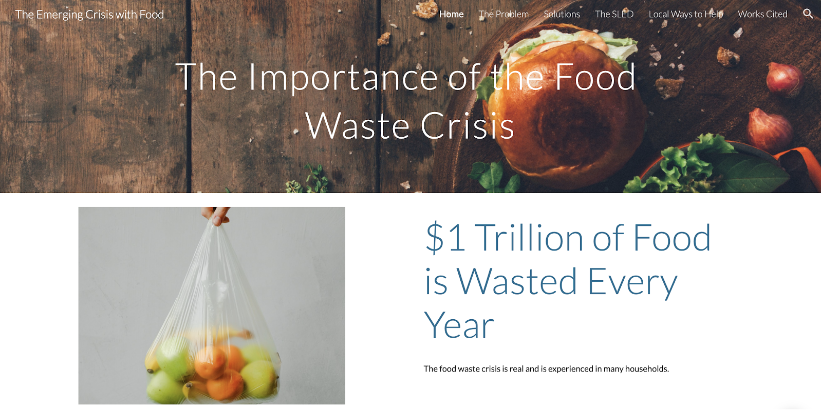}
\end{center}

\centering \emph{Figure 12: A website was created to help increase awareness about the food waste crisis. }

\justifying

\vspace{5mm} %5mm vertical space

\subsubsection{Future Work}

Next steps for this project include a feature where consumers can find local food drives near them to donate food that they may not eat in time and may potentially be wasted. Reducing the cost of the test kit will allow it to be more accessible and scalable. This will be done by leveraging automation in addition to sourcing with cheaper materials while preserving quality. One current limitation of the app is that only the five tested foods can be tracked by consumers. Adding more foods to the database allows consumers to test a greater variety of foods for spoilage. Additional trials on current foods will also provide a greater number of data points that will generate a more accurate algorithm. 

\section{CONCLUSIONS}

This experiment aimed to develop a correlation between the amount of time until a food spoiled and dates on food labels in conjunction with sensory observations. The hypothesis that 1) the actual time until spoilage would be longer than that of the best by date and 2) sensory observations would prove to be the most accurate factor in determining spoilage were both supported. All five foods tested lasted significantly longer than the date printed on the food label. Sensory observations were able to detect spoilage accurately, proven by the strong correlations with the day these observations were taken from the regression graphs.

From this data, an app and test kit were created to create a system in which the consumer could control their food. The app contains features where the consumer can input observations and find how long their food is expected to last until spoilage. It also allows consumers to monitor their food by alerting them to check on their food if it may spoil soon. The test kit allows the user to test the characteristics themselves, in addition to serving as an alternative for people without access to the technology the app runs on. 

The Shelf Life Expiration Date (SLED) Tracking System is a viable, consumer centered solution that proves that the spoilage of foods is quantifiable and be mathematically modeled over time to not only prevent food-induced illness, but also prevent food waste as well. 

\vspace{3mm} %3mm vertical space

\textbf{Conflict of Interest}: The author has declared that there is no conflict of interest. 

\vspace{3mm} %3mm vertical space

\textbf{Data Availability Statement}: All observational data  can be provided upon request. 

\section{REFERENCES}

\RaggedRight

\vspace{3mm} %3mm vertical space

[1] Jacobson, Kurt. “The Environmental Impact of Food Waste.” Move For Hunger, 2015, moveforhunger.org/the-environmental-  impact-of-food-waste.
 
[2] Karoui, Romdhane. “Methods to Evaluate Egg Freshness in Research and Industry: A Review.” European Food Research and   Technology, Mar. 2006, pp. 727–732.,   doi:10.1007/s00217-005-0145-4.

[3] Lineberry, Kristopher R, et al. “A Proposed Method of Test for Spoilage of Fruits and Vegetables.” Iowa University Food Science and Human Nutrition Publications, vol. 5, 2012. 

[4] Lu et al. “Milk Spoilage: Methods and Practices of Detecting Milk Quality.” Food and Nutrition Sciences, July 2013, pp. 113–  123., doi:10.4236/fns.2013.47A014.  

[5] Sanchez-Gonzalez, Jesus Alexander, and Jimy Frank Oblitas-Cruz. “Application of Weibull Analysis and Artificial Neural Networks to Predict the Useful Life of the Vacuum Packed Soft Cheese.” Revista Facultad De Ingeniería, Universidad De Antioquia, vol. 82, 2017, pp. 53–59. 

[6] Steinberg, Lynn. “Family, Parenting, Pet and Lifestyle Tips That Bring Us Closer Together.” LittleThingscom, 2017, littlethings.com/home/how-to-tell-if-eggs-are-bad. 

[7] Stojiljković, Mirko. “Linear Regression in Python.” Real Python, Real Python, 26 Nov. 2020, realpython.com/linear-regression-in-python/. 

[8] “This Is The Equally Harmful Flipside of Food Waste: Food Loss.” World Food Program USA, 11 Nov. 2020m, www.wfpusa.org/explore/wfps-work/drivers-of-hunger/foodwaste/. 

[9] wikiHow. “How to Tell If Milk Is Bad.” WikiHow, WikiHow, 6 Feb. 2021, www.wikihow.com/Tell-if-Milk-is-Bad. 

\end{document}